\documentstyle{mn}

\def\pmb#1{\setbox0=\hbox{#1}%
    \kern-.025em\copy0\kern-\wd0
    \kern.05em\copy0\kern-\wd0
    \kern-.025em\raise.0433em\box0 }
\def\btheta{\pmb{$\theta$}}

\newcounter{parentequation}

\def\ltsima{$\; \buildrel < \over \sim \;$}
\def\gtsima{$\; \buildrel > \over \sim \;$}
\def\simlt{\lower.5ex\hbox{\ltsima}}
\def\simgt{\lower.5ex\hbox{\gtsima}}

\def\etal{{\it et al.}\rm}
\def\etals{{\it et al. }\rm}

\begin{document}

\title[Sunyaev-Zeldovich Template]
{A Simple Empirically Motivated Template for the Unresolved Thermal Sunyaev-Zeldovich  Effect}

\author[G. Efstathiou \& M. Migliaccio]{George Efstathiou$^{1}$ and Marina Migliaccio$^{1,2}$\\
1. Kavli Institute for Cosmology and Institute of Astronomy, Madingley Road, Cambridge, CB3 OHA.\\
2. Dipartimento di Fisica dell'Universit\`a di Roma ``Tor Vergata'', Via della
Ricerca Scientifica 1 00113,  Roma, Italy.}

\maketitle

\begin{abstract}
  We develop a model for the power spectrum of unresolved
  clusters of galaxies arising from the thermal Sunyaev-Zeldovich
  (tSZ) effect. The model is based on a `universal' gas pressure
  profile constrained by X-ray observations and includes a parameter
  to describe departures from self-similar evolution.  The model is
  consistent with recent Planck observations of the tSZ effect for
  X-ray   clusters with redshifts $z \simlt 1$ and reproduces the low
  amplitude for the tSZ inferred from recent ground based
  observations. By adjusting two free parameters, we are able to
  reproduce the tSZ power spectra from recent numerical simulations to
  an accuracy that is well within theoretical uncertainties. Our
model provides a simple, empirically motivated tSZ template that
may be useful for the analysis of new experiments such as Planck.

\vskip 0.15 truein


\vskip 0.1 truein

\end{abstract}

\section{Introduction}

The thermal Sunyaev-Zeldovich signal \cite{Sunyaev72}, caused by
inverse Compton scattering of cosmic microwave background (CMB)
photons by the hot plasma in clusters of galaxies, has been detected
convincingly by many experiments (see Carlstrom, Holder and Reese,
2002, for a review). It has long been recognised that the integrated
tSZ signal from distant, faint, unresolved clusters of galaxies would
make a significant `secondary' frequency-dependent contribution to the
CMB temperature power spectrum at high multipoles
\cite{CK98}. However, there are many other contributors to the
anisotropies at multipoles $\ell \simgt 2000$, principally Poisson
radio sources at low frequencies ($\nu \simlt 100 {\rm GHz}$),
clustered and unclustered infra-red galaxies at higher frequencies,
together with the frequency-independent secondary anisotropies
associated with cluster peculiar motions and inhomogeneous
reionization (see {\it e.g.} Iliev \etals 2007). Isolating the
unresolved tSZ contribution requires disentangling these various
contributions.

This has become possible recently using high resolution observations
of the CMB by the Atacama Cosmology Telescope (ACT, Dunkley \etals
2010) and by the South Pole Telescope (SPT, Lueker \etals 2010). Both
the ACT and SPT teams perform multi-parameter fits to the temperature
power spectra using `templates' to model the secondary anisotropies.
They find statistically significant evidence for a tSZ contribution,
but with an amplitude at a frequency of $\approx 150\ {\rm GHz}$ of
only a few $(\mu {\rm K})^2$ ($4.2 \pm 1.5 (\mu {\rm K})^2$ at
$\ell=3000$ for SPT, and $6.8 \pm 2.9 (\mu {\rm K})^2$ from ACT for
the combined thermal and kinetic SZ effects). These amplitudes are
significantly smaller than expected from semi-analytic predictions
using the WMAP5 parameters (see {\it e.g.}  Komatsu and Seljack 2002).

The earliest approaches to computing the unresolved tSZ contribution
involved adopting a model for the pressure profiles of clusters
combined with a Press-Schechter \cite{PS74} type theory to compute
their spatial abundance as a function of mass and redshift (Cole and
Kaiser 1988; Bond and Myers 1996; Cooray 2000, 2001; Komatsu and
Seljack 2002). These calculations established the strong sensitivity
of the unresolved tSZ amplitude to the normalization of the spectrum
of fluctuations (roughly varying as $\sigma_8^7$, where $\sigma_8$ is
the {rms} fluctuation amplitude at the present day in spheres of
radius $8\ h^{-1}{\rm Mpc}$\footnote{$h$ is the Hubble constant in
  units of $100\ {\rm km}{\rm s}^{-1}{\rm Mpc}^{-1}$.}). This led to
the hope that observations of the unresolved tSZ effect could be used
to constrain the amplitude of scalar fluctuations.

An alternative approach to modeling the tSZ effect is to use numerical
hydrodynamic simulations incorporating as much realistic physics as
possible ({\it e.g.}  de Silva \etals 2000; Springel, White and
Hernquist 2001; Bond \etals 2005: Lau, Kratsov and Nagai, 2009;
Battaglia \etals 2010).  Various other approaches have been used,
including dark matter simulations (Bode, Ostriker and Vikhlinin 2009;
Sehgal \etals 2010; Trac, Bode and Ostriker 2010) or Press-Schechter
type calculations (Shaw \etals 2010) combined with semi-analytic
prescriptions for assigning pressure profiles to dark matter halos
incorporating schematic models for star formation and feedback from
supernovae and active galactic nuclei.

\begin{figure*}

\vskip 2.9 truein

\includegraphics{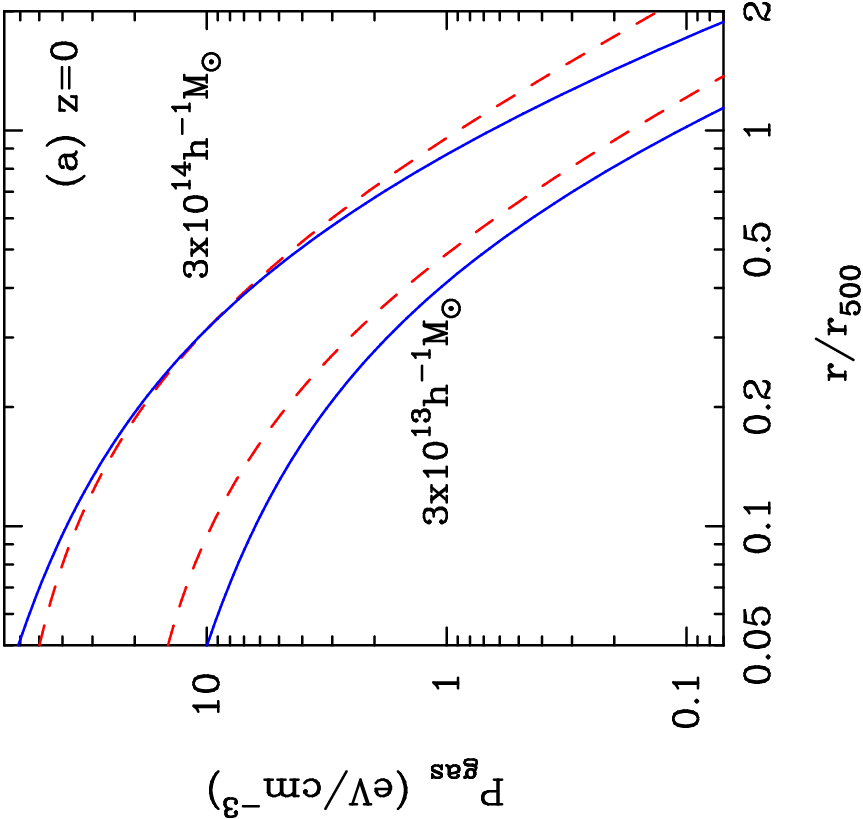}
\includegraphics{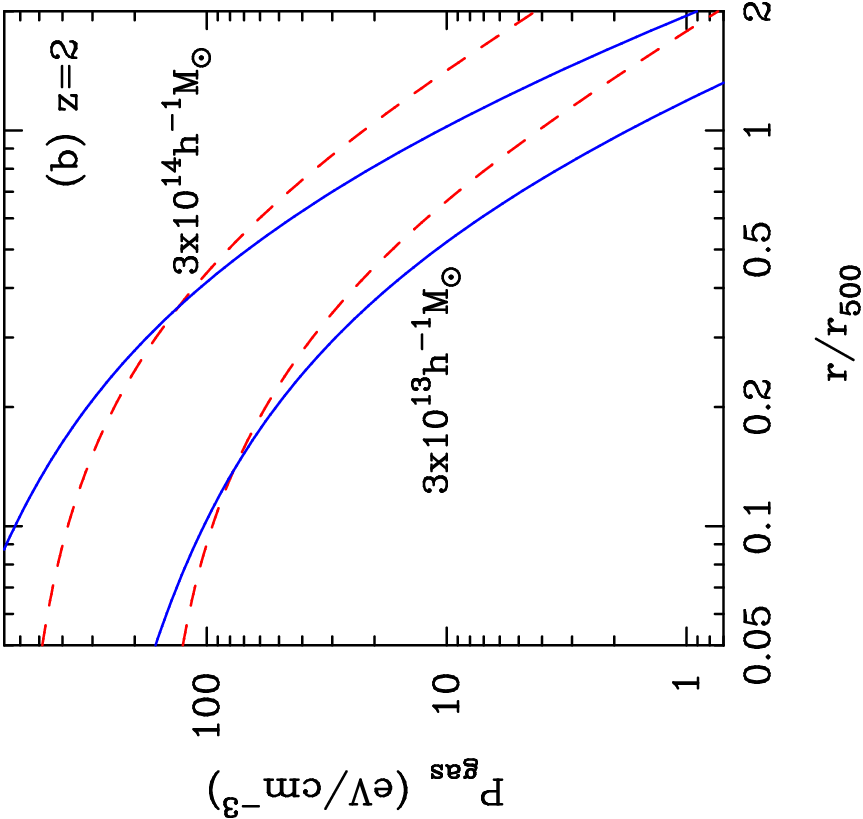}

\caption
{Gas pressure profiles for clusters with masses $M_{500}=3 \times
  10^{14}h^{-1} M_\odot$ and $3 \times 10^{13}h^{-1} M_\odot$ at $z=0$ (a)
  and $z=2$ (b). The solid lines show the X-ray `universal' pressure
  profile of equation (\ref{P1}) and the dashed lines show the
  analytic profile of Komatsu and Seljak (2002) computed for the
  cosmological parameters adopted in this paper and the revised halo
  concentration parameter of equation (\ref{C1}). The X-ray profile in
  Figure 1(b) is plotted for self-similar evolution, {\it i.e.}
  $\epsilon=0$ in equation (\ref{P1}).}
\label{figure1}
\end{figure*}

The numerical hydrodynamic simulations, in particular, have shown just
how sensitive cluster pressure profiles are to complex physics. Within
$\sim 0.2\ r_{500}$ ($r_{500}$ is the radius at which the cluster has
a mean overdensity of $500$ times the critical density at the redshift
of the cluster), the pressure profiles are sensitive to the
prescriptions for star formation and feedback. At larger radii, the
pressure profiles differ from those expected from hydrostatic
equilibrium because of the increasing importance of non-thermal
motions. Although there has been remarkable progress in the
sophistication of numerical hydrodynamic simulations, the physics
involved is complex and this is reflected in the relatively large
scatter between predictions of the unresolved tSZ power spectrum (see {\it
e.g} Fig 3 of Battaglia \etals 2010). Early expectations 
that measurements of the  tSZ effect (in particular, number
counts and the power spectrum) could be used for
precision cosmology now seem naive. It is more likely
that such measurements will provide constraints on the complex
physics that structures the intra-cluster medium.

Since the unresolved thermal SZ effect is constrained by fitting a
template to the observed power spectra, how should the template be
chosen?  Should experimentalists adopt one or more highly model specific templates
determined from hydrodynamic simulations? Should the uncertainties
in the physics be represented by a large number of adjustable 
parameters? Or should experimentalists adopt  a phenomenological
model with fewer parameters that may be less closely linked
to the physics.

In this short paper, we adopt an empirical approach to computing the
tSZ power spectrum. The model is based on the \cite{KS02} model (with
minor modifications) but instead of using the theoretical pressure
profiles computed by Komatsu and Seljak (2001) we use the `universal'
pressure profiles derived from X-ray observations \cite{Arnaud10}. 
Moreover, we introduce an additional parameter, $\epsilon$,  to model
deviations from self-similar evolution of the cluster profiles. The
resulting model is simple, empirically motivated, and provides a
flexible tSZ template that can match the results from recent numerical
simulations.

\section{The model}

Unless otherwise stated, we adopt the cosmological parameters from the
6 parameter $\Lambda CDM$ model from Table 3 of \cite{Komatsu11},
namely $h=0.71$, $\sigma_8 = 0.80$, $n_s = 0.963$, $\Omega_\Lambda =
0.734$, $\Omega_b=0.0448$\footnote{The additional parameter, the
  optical depth $\tau$ from late reionization of the inter-galactic
  medium, is unimportant for this study.}. We assume a spatially flat
Universe, $\Omega_k=0$, and assume that the dark energy is a
cosmological constant with equation of state $p=-\rho c^2$.

For a  Poisson distribution of clusters
of mass $M$ and comoving number mass-function $dn/dM$, 
 the power spectrum of the tSZ effect is given by
\begin{equation}
C_\ell =  g^2(\nu) T_0^2\int_{z_{\rm min}}^{z_{\rm max}} dz {dV \over dz} \int_{M_{\rm min}}^{M_{\rm max}} {d n 
\over dM} \vert \tilde y_\ell (M, z) \vert^2 dM, \label{SZ1}
\end{equation}
(Komatsu and Seljak 2002, hereafter KS02). Here $g(\nu)$ 
describes the spectral dependence of the tSZ effect, 
which in the non-relativistic limit is given by
\begin{equation}
g(\nu) = \left( x{e^x + 1 \over e^x-1} - 4 \right ), \quad x={h_p \nu \over kT_0}, \label{SZ2}
\end{equation}
where $T_0$ is the present temperature of the CMB and $h_p$ is
Planck's constant.  We will show results for a frequency of $\nu = 143
\ {\rm GHz}$ corresponding to Planck's most sensitive channel for
detection of the tSZ effect (Planck Collaboration 2011a, b, c) and close to the
frequencies of the tSZ sensitive channels of ACT ($148 \ {\rm GHz}$)
and SPT ($150 \ {\rm GHz}$). The remaining terms in (\ref{SZ2}) are as
defined in KS02. In particular, $\tilde y_\ell$ is the two-dimensional
Fourier transform of the Compton y parameter:
\begin{equation}
\tilde y_\ell =  4 \pi {r_{500} \over \ell^2_{500}} \int_0^\infty dx x^2 Y_{3D}(x) 
{\sin(\ell x/\ell_{500}) \over (\ell x/\ell_{500})},  \label{SZ3}
\end{equation}

\begin{center}
Table 1
\begin{tabular}{llllll} \hline
              & $P_0$ & $c_{500}$ & $\gamma$ & $\alpha$ & $\beta$ \\ \hline
All           & $4.921$ & $1.177$ & $0.3081$ & $1.0510$ & $5.4905$  \\
Cool core     & $1.902$ & $1.128$ & $0.7736$ & $1.2223$ & $5.4905$  \\
Non-cool core & $1.875$ & $1.083$ & $0.3798$ & $1.4063$ & $5.4096$  \cr \hline
\end{tabular}
\end{center}

\medskip

\noindent
where $Y_{3D}$ is the three-dimensional Compton y-profile,
$$
 x \equiv {r \over r_{500}}, \quad \ell_{500} \equiv {d_A(z) \over r_{500}}, 
$$
and $d_A(z)$ is the angular diameter distance to a cluster at redshift
$z$.  Notice that we have used $r_{500}$ as a characteristic radius,
rather than the scale radius $r_s$ of the dark matter distribution
used by KS02.

The three-dimensional Compton profile is given by
\begin{equation}
Y_{3D}(x) = {\sigma_T  \over m_ec^2} P_e (x) = 
{\sigma_T  \over m_ec^2} \left ( {2 + X \over 3 + 5X} \right )P_{\rm gas} (x) \label{SZ4}
\end{equation}
where $X=0.76$ is the primordial Hydrogen abundance and $P_e$ and $P_{\rm gas}(x)$ are the electron
and gas pressure profiles.  X-ray data of the REXCESS cluster sample \cite{Arnaud10} suggest that clusters are
well described by a `universal' electron pressure profile of the form:
\begin{equation}
P_{\rm e}(x) = 1.88\left [{ M_{500} \over 10^{14}h^{-1} M_\odot} \right ]^{0.787} p(x) E(z)^{{8\over 3} - \epsilon}h^2 {\rm eV}
\ {\rm cm}^{-3}, \label{P1}
\end{equation}
where 
\begin{equation}
p(x) ={ P_0 h^{-3/2}  \over (c_{500} x)^{\gamma} (1 + [c_{500}x]^{\alpha})^{(\beta - \gamma)/\alpha}},  \label{P2}
\end{equation}
with the parameters given in the first row of Table 1. The function $E(z)$ in (\ref{P1}) is the ratio
of the Hubble parameter at redshift $z$ to its present value,
$$
E(z) = \left[ (1 - \Omega_\Lambda) (1+z)^3 + \Omega_\Lambda \right]^{1/2}, 
$$
and the scaling $E(z)^{8/3}$  in (\ref{P1}) 
is appropriate for self-similar evolution. The parameter $\epsilon$
 therefore describes departures from self-similar evolution.

The profile (\ref{P1}) is constrained from X-ray observations out to
radii $r \sim r_{500}$ but the extrapolation beyond $r_{500}$ was
designed to fit results from numerical simulations of relaxed clusters
\cite{NVK07}. Since a significant fraction of the tSZ signal comes
from $r > r_{500}$, it is important to recognise that an unresolved
tSZ template based on (\ref{P1}), though empirically motivated, relies
on : (a) an extrapolation of the pressure profiles beyond the
observed range of radii; (b) a highly uncertain extrapolation of the
shapes of the profiles to high redshift; (c) the assumption that well
observed X-ray clusters are representative of the cluster population
as a whole. These points will be discussed in further detail below.

Reliable measurements of the gas temperature in the faint cluster
outskirts have become available only recently, allowing the
characterization of the pressure behavior of a few clusters out to the
virial radius (approximately $2r_{500}$) or beyond.  Examples are the Perseus
Cluster \cite{Simionescu11}, a massive and relaxed cluster observed
with the \textit{Suzaku} satellite , and the dynamically
younger and lower mass Virgo Cluster
observed with \textit{XMM-Newton} \cite{Urban11}.  After correcting for clumping of
the gas at large radii in Perseus, both clusters show a pressure
profile  in good agreement with (\ref{P1}) out to $\sim 2r_{500}$. 
Further evidence in favour of the `universal' pressure profile 
beyond $r_{500}$ comes from the stacked radial 
 SZ profiles of 15 X-ray selected clusters observed with SPT
\cite{Plagge10}.  

A recent analysis by Sun \etals (2011) of nearby ($z<0.12$) galaxy
groups observed with \textit{Chandra} shows that their their median
pressure profile is also well described by (\ref{P1}) out to $\sim
r_{500}$. However, little is known about the pressure profiles of
groups at larger radii, or at higher redshifts.

Figure 1 compares the X-ray inferred gas pressure profile of equation
(\ref{P1}) with the analytic pressure profiles used by KS02 for
clusters with masses $M_{500}=3 \times 10^{13}h^{-1}M_\odot$ and
$M_{500}= 3\times 10^{14} h^{-1} M_\odot$ at $z=0$ and $z=2$ assuming
$\epsilon=0$. To compute the Komatsu-Seljak profiles we have assumed a
Navarro, Frenk and White (1997) dark matter profile,
\begin{equation}
\rho_H = {\rho_0 \over (r/r_s)(1 + r/r_s)^2},  \label{P4}
\end{equation}
but with the revised concentration parameter relating the scale radius to the virial radius
($r_s = r_V/c$) derived by \cite{Duffy08}
\vskip 0.1 truein
\begin{equation}
c(M_V, z) = 5.72 \left ( {M_V \over 10^{14}h^{-1} M_\odot} \right)^{-0.081} (1+z)^{-0.71}. \label{C1}
\end{equation}
The definitions of the virial radius and virial mass used here follow
those of KS02. The masses $M_{500}$ inferred from X-ray observations
assume hydrostatic equilibrium. Numerical simulations ({\it e.g.}
Nagai, Vikhlinin and Kravtsov 2007; Lau, Kravtsov and Nagai 2009;
Battaglia \etals 2010; Nagai 2011) show that non-thermal pressure
becomes significant by $r_{500}$ and that assuming hydrostatic
equilibrium underestimates the true mass $M_{500}$ by about $10\%$. We
have therefore assumed a $10\%$ correction factor to relate the X-ray
mass to the true mass.

\begin{figure}

\vskip 2.9 truein

\includegraphics{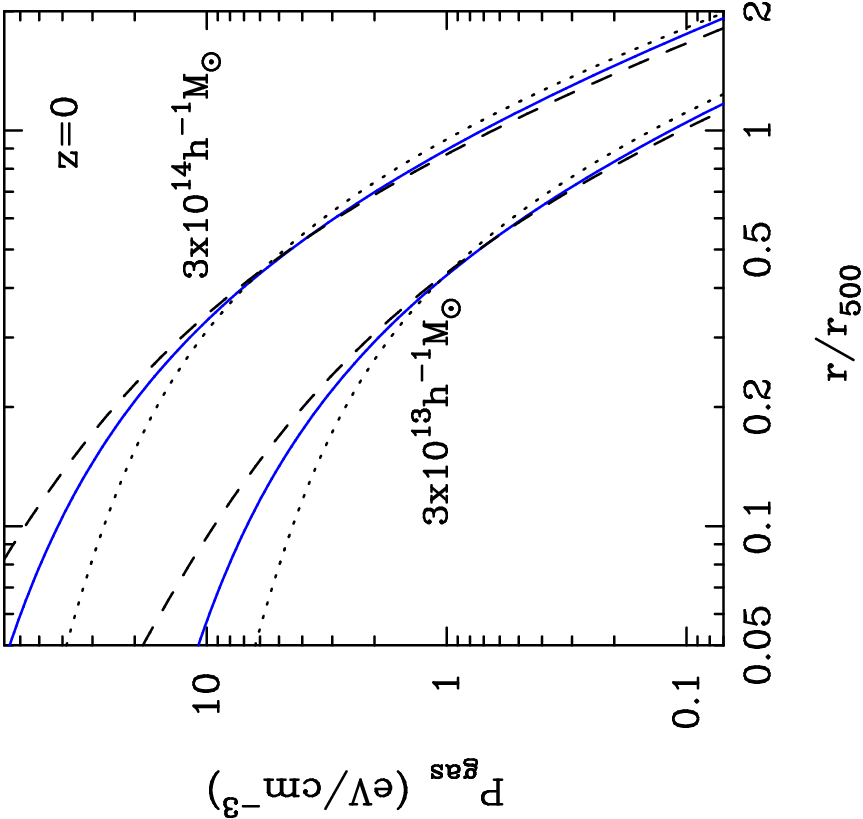}

\caption
{Comparison of the pressure profiles with parameters listed in Table 1.
Solid (blue) lines show the average profile of all REXCESS clusters
(first row of Table 1). The dashed and dotted line shows the average profile
for cool core and non-cool core clusters respectively
(second and third rows of Table 1).}

\label{figure2}
\end{figure}

At radii $r \simgt 0.3 r_{500}$ (most relevant for the unresolved tSZ
effect) the X-ray inferred pressure profiles lie below the KS02
profiles (compare Figure 17 of Komatsu \etals 2011). This is 
expected because the KS02 profile is derived assuming an 
equation of state with a constant polytropic index and 
does not account for the increasing importance of non-thermal
pressure at large radii. In fact, the integrated Compton Y-parameter
$\int Y_{3D} x^2 dx$ does not converge for the KS02 profiles. As in 
KS02, we arbitrarily truncate the integrals (\ref{SZ3}) at $r = 2r_V$ to
compute the power spectrum. (In contrast, the Compton Y-parameter for 
the X-ray pressure profiles converges and we adopt an upper cut-off
of $4r_V$ for these profiles).

\begin{figure*}

\vskip 2.6 truein

\includegraphics{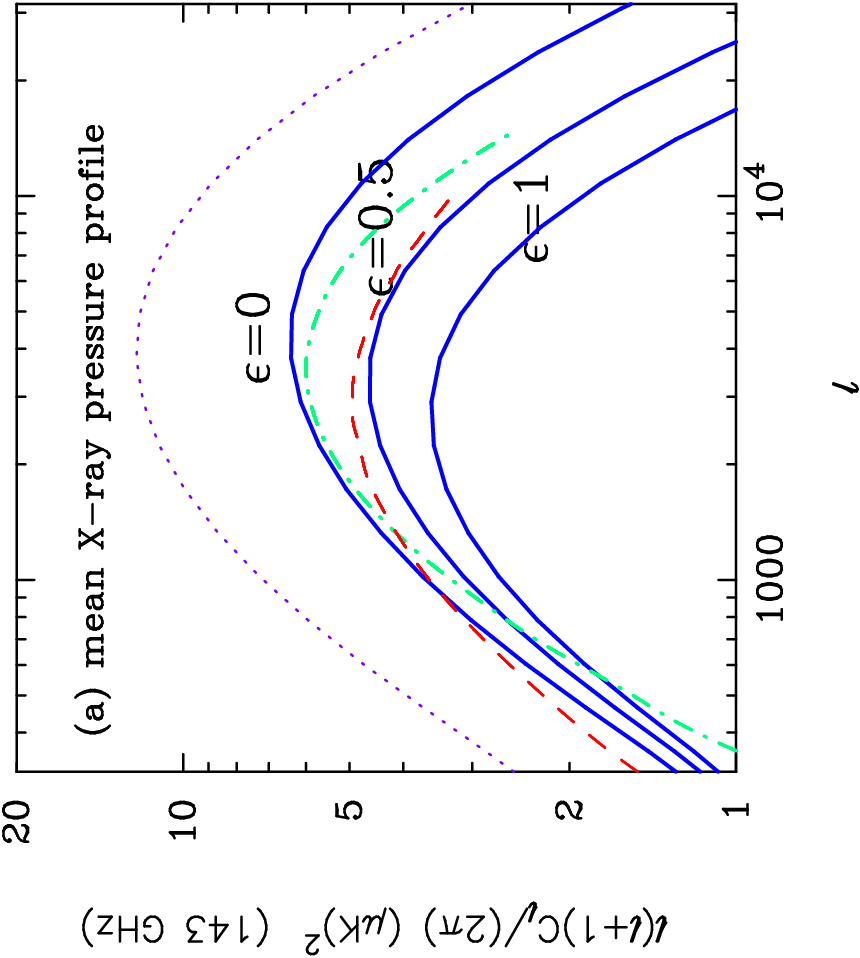}
\includegraphics{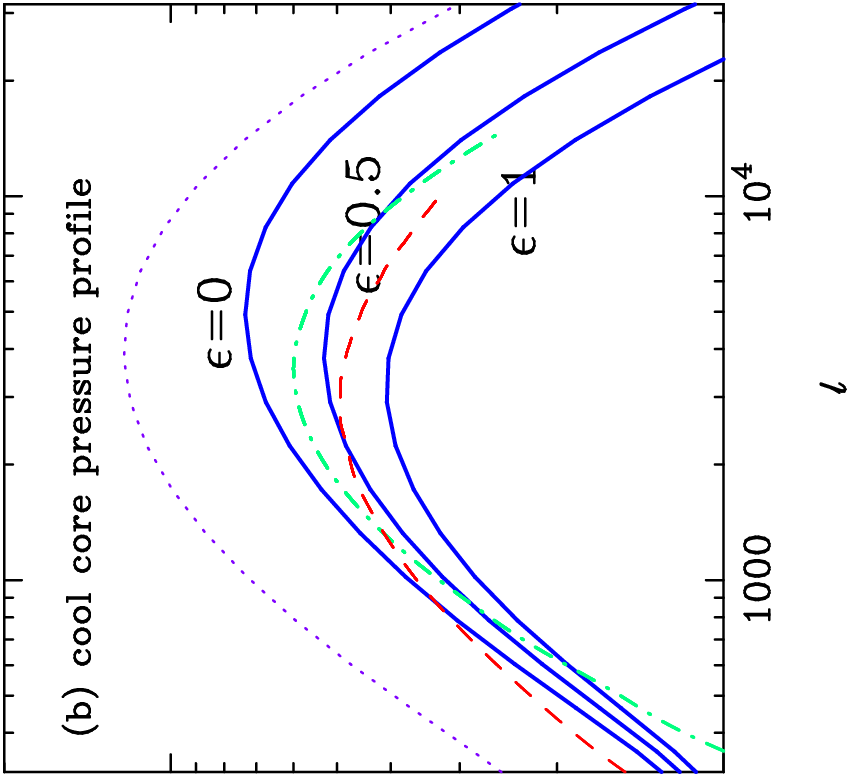}
\includegraphics{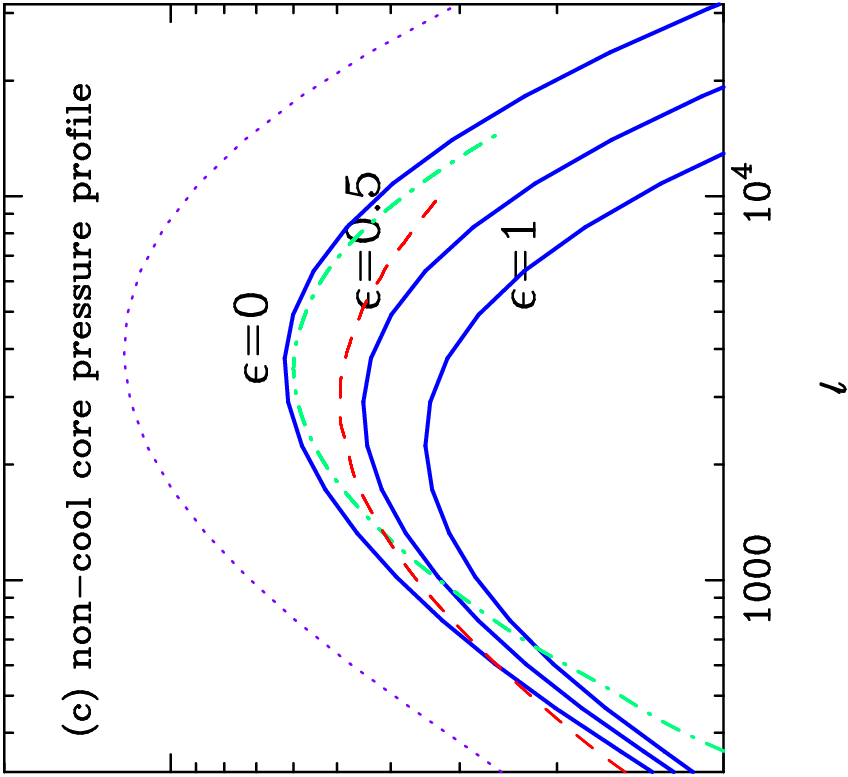}

\caption
{Predictions for the unresolved tSZ power spectrum. The dotted (purple) lines show the KS02
model computed using the WMAP7 cosmological parameters and the concentration parameter
of equation (\ref{C1}). The solid lines (blue) show computations 
using the X-ray pressure profile of equation (\ref{P2}) for three values of the
evolution parameter $\epsilon$ : panel (a) shows the mean
profile for all REXCESS clusters; panel (b) for cool core clusters; panel (c)
for non-cool core clusters. The (green) dot-dashed line and red (long dashed) 
line in each panel show two templates used by the ACT team (Dunkley \etals 2010):
dot-dashed line shows the  template from the AGN feedback simulations of 
Battaglia \etal (2011); long-dashed line shows the TBO-2 template from the
numerical simulations of Trac, Bode and Ostriker (2011).}
\label{figure3}
\end{figure*}

As mentioned above, the profile (\ref{P2}) is adjusted to match the
numerical results of Nagai, Vikhlinin and Kravtsov (2007) at radii $r
\simgt r_{500}$. The pressure profiles from the AGN feedback
simulations of Battaglia \etals (2010) at $z=0$ fall off slightly less
rapidly than equation (\ref{P2}) at $r \simgt r_{500}$. However, they
find that the outer pressure profiles of clusters at a redshift $z = 1$
(which make the dominant contribution to $C_\ell$ at $\ell \sim 3000$)
are steeper and in reasonable agreement with equation (\ref{P2}).

The X-ray inferred pressure profiles for cool core clusters differ
systematically from those of non-cool core (often morphologically
disturbed) clusters, sometimes differing by more than an order of
magnitude at $r \simlt 0.2r_{500}$ (Arnaud \etals 2010). However,
there is no evidence for systematic differences in the pressure
profiles at larger radii. In fact, it is the pressure profiles at $r
\simgt 0.2r_{500}$ that dominate the power spectrum. The inner
pressure profiles have a relatively small effect on the shape of the
power spectrum at high multipoles. To illustrate this, we have
computed power spectra using fits of equation (\ref{P2}) to the mean
pressure profiles of cool core and non-cool core clusters
\cite{Planck11d}. The parameters for these fits are listed in Table 1
and the pressure profiles are plotted in Figure 2.

Computations of the tSZ power spectrum for a frequency of $143\;{\rm
  GHz}$ are shown in Figure 3.  The dotted (purple) lines in each
panel show the KS02 model.  As in KS02 we used the Jenkins \etals
(2001) mass function in equation (1) and fixed other parameters ({\it
  e.g.}  $M_{\rm min}$, $z_{\rm max}$) to those used in KS02.  The
power spectra plotted in Figure 3 therefore differ from those of KS02
only because of our choice of cosmological parameters and
concentration relation $c(M_V, z)$. The peak amplitude of this model
is about $12 \ (\mu {\rm K})^2$, i.e. about three times higher than the
amplitude inferred from ACT and SPT (Dunkley \etals 2010; Lueker
\etals 2010).

The predictions of our model are shown by the solid lines in each
panel for three values of the evolution parameter $\epsilon$. The
three panels show the sensitivity of the models to the shape of the
inner pressure profiles.  These are relatively minor, except at
multipoles $\ell \simgt 10^4$, which are extremely difficult to probe
experimentally. Significantly, the peak amplitude of these models is
in the range $3$ -- $6\; (\mu {\rm K})^2$, consistent with the 
constraints from ACT and SPT. {\it The observations are therefore
consistent with a model based on the X-ray `universal' pressure
profile and nearly self-similar evolution.}

All of the models in Figure 3 assume a fluctuation amplitude of
$\sigma_8=0.8$. The power spectra of our models scale with amplitude
as $C_\ell \propto \sigma_8^{7.9}$ and so variations of, say, $10\%$
in $\sigma_8$ have a much greater effect than the variations in the
shapes of the pressure profiles and $\epsilon$ explored in Figure 3.
Note that if the KS02 model is used to fit the observations, the
inferred value of $\sigma_8$ would be underestimated by about $14\%$.

The statistical cross correlation of Planck maps with X-ray detected
clusters\footnote{The Meta-Catalogue of X-ray detected Clusters
  \cite{P11}, supplemented with $33$ clusters at redshifts $z>0.6$.}
provides a constraint on the evolution parameter $\epsilon$
\cite{Planck11d}. These authors applied a multifrequency matched
filter algorithm \cite{Melin06} to the Planck data 
at the positions of the X-ray clusters to determine an
integrated  Compton parameter $Y_{500}$. The Planck
data follow a relation
\begin{equation}
{d^2_A(z) Y_{500} \over M^{1.783}_{500}} \propto E(z)^{2/3 + \epsilon},
\end{equation}
with $\epsilon = 0.66 \pm 0.52$ for redshifts $z < 1$ (see Figure 6 of
Planck Collaboration 2011d). These observations do not extend to the
very high redshifts $z \sim 3$ that contribute to the tSZ power
spectrum at multipoles $\ell \sim 10^4$, but they provide partial
overlap with the redshift range contributing at lower multipoles. The
observations are consistent with self-similar evolution, $\epsilon=0$,
with perhaps some indication of weaker evolution. Extending this type
of analysis to higher redshifts would clearly help in developing
models of the unresolved tSZ effect.

\begin{figure*}

\vskip 2.9 truein

\includegraphics{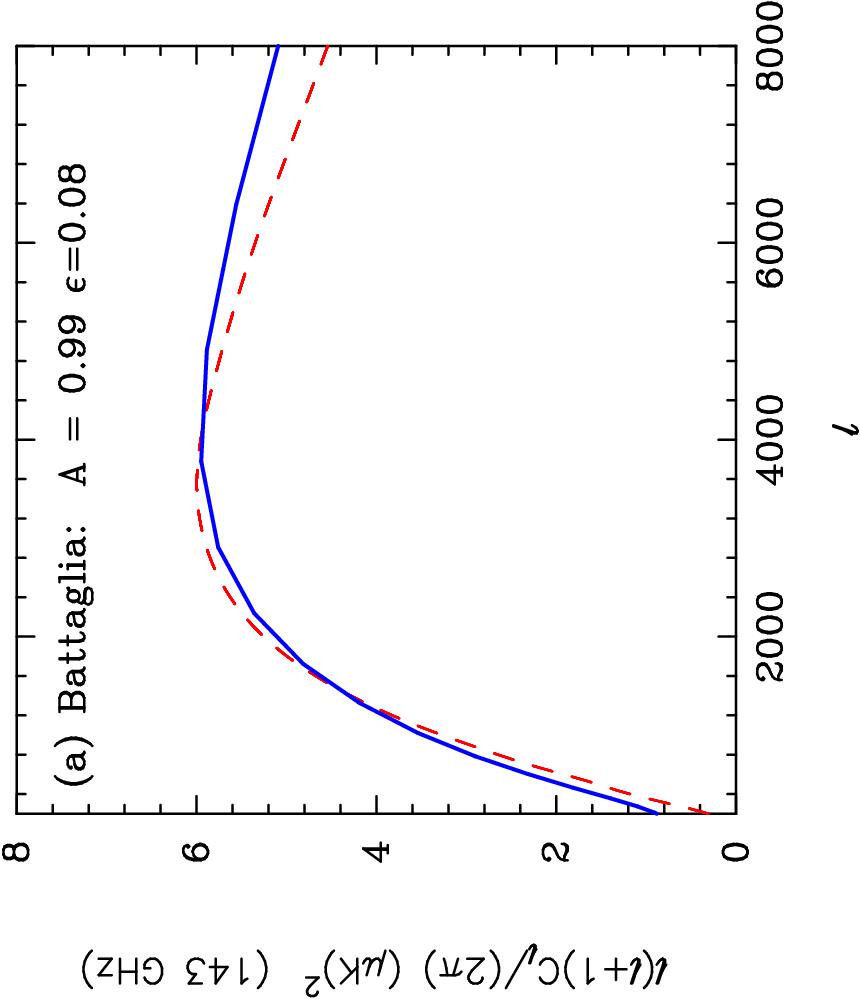}
\includegraphics{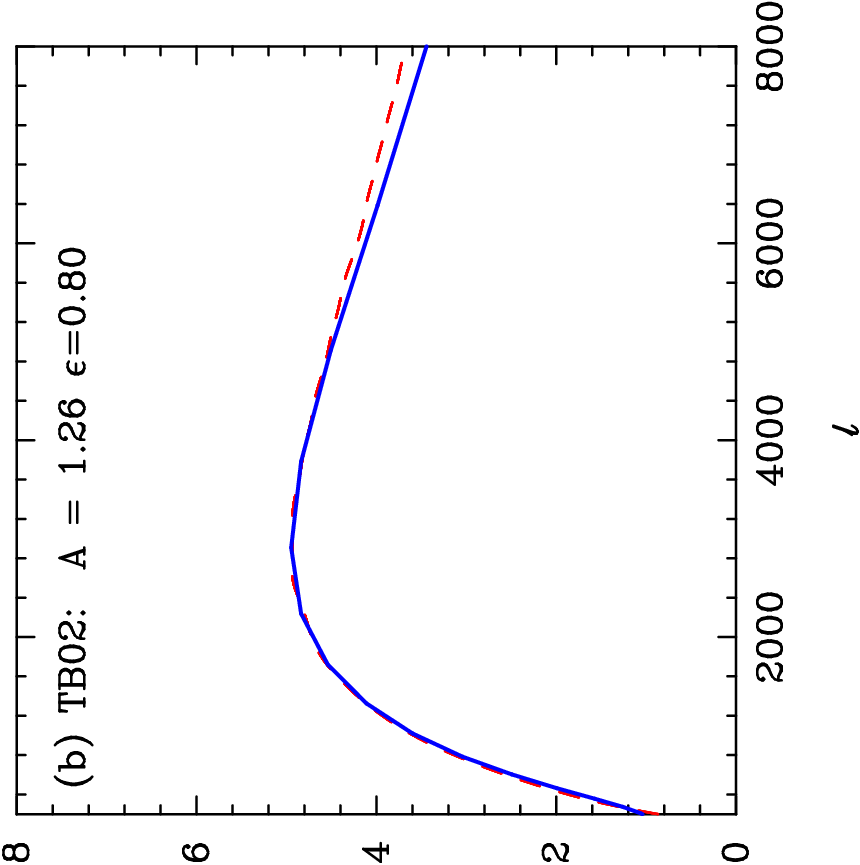}

\caption {Fits of our template model (blue/solid lines) to the 
simulation  templates (red/dashed lines). Panel (a) shows the Battaglia 
template and panel (b) shows the TBO-2 template. The best fit parameters
$A$ and $\epsilon$ are given in each panel.}

\label{figure4}
\end{figure*}

The dashed and dot-dashed lines in Figure 3 show two of the tSZ
templates used in the ACT analysis. The (red) dashed line shows the
AGN feedback template of Battaglia \etals (2011), while the
(green) dot-dashed line shows the `nonthermal20' model of
 Trac, Bode and Ostriker (2011). Following Dunkley
\etals (2010) we will refer to the former as the `Battaglia' template
and the latter as the `TBO-2' template. These templates are based on
very different approaches. The Battaglia template is derived from
hydrodynamic simulations, while the TBO-2 template is based on
post-processing dark matter simulations by assigning pressure profiles
to dark matter halos.  These templates give some indication of the
theoretical uncertanties involved in computing the tSZ power spectrum.
It is encouraging that both templates have about the same peak
amplitude, consistent with the observations, but the slopes at 
both low and high multipoles differ. (Note that because of the
finite computational volumes the variance of these
templates at low multipoles is high and not accurately quantified).

The physical processes involved in determining the shape and amplitude
of the tSZ power spectrum are complicated, and it will not be easy to
improve the accuracy of the simulation templates.  As mentioned in the
Introduction, the choice faced by experimentalists is either to use a
number of templates spanning the range of theoretical uncertainties
(with possible uncertain scalings with cosmological parameters) or to
adopt a parametric model. In our approach, the parametric model is
empirically motivated and has a small number of free parameters
(principally the amplitude and the evolutionary parameter
$\epsilon$\footnote{Other cosmological parameters, such as
  $\Omega_\Lambda$ and $\Omega_k$ are now so well constrained that
  their errors can be ignored.}). Nevertheless, to be useful, our
model should have sufficient flexibility to match the simulation
templates. This is illustrated in Figure 4.  For each simulation
template, $C_\ell^{\rm SZsim}$, we find the amplitude $A$ and
$\epsilon$ parameter that minimises
\begin{equation}
\chi^2 = \sum_\ell [C_\ell^{\rm SZsim} - A C_\ell(\epsilon)]^2,
\end{equation}
where the sum extends over the range $1000 \le \ell \le 6000$. As can
be seen from Figure 4, our model matches the Battaglia template to
an accuracy of better than $10\%$ over the multipole range plotted 
in the figure and matches the TBO-2 template to even higher accuracy.
These errors are considerably smaller than the theoretical
uncertainties in the simulation templates. For the Battaglia
template, the best fit amplitude is $A=0.99$, so the use of
our model would not bias a measurement of $\sigma_8$. The best
fit amplitude for the TBO-2 template is $A=1.26$. If the
TBO-2 template were correct, using our model would lead
to a downward bias of $3\%$ in a measurement of $\sigma_8$.

One key point, that is not yet well understood, is whether the
pressure profiles of X-ray selected clusters are representative of the
cluster population as a whole. As Arnaud \etals (2010) have stressed,
although the X-ray luminosities of non-cool core clusters differ
systematically from those of cool core clusters of the same mass,
their pressure profiles at $r \simgt 0.3 R_{500}$ are almost identical
({\it c.f.}  Figure 2). However, a cross-correlation of Planck maps
with rich clusters selected from the Sloan Digital Sky Survey (Planck
Collaboration 2011e) has revealed a possible discrepancy with X-ray
`universal' pressure profile. The observed correlation between
$Y_{500}$ and optical richness,  $N_{200}$,  lies below the X-ray model by
a factor of $\sim 1.7$ and $\sim 2.2$,  depending on which empirical 
weak-lensing mass calibration is used to convert $N_{200}$
to mass. It is not yet clear whether this result is indicative of a
population of sub-luminous tSZ clusters under-represented in X-ray
surveys, whether it is caused by some systematic error in the weak
lensing mass estimates or some other systematic error such as optical
projection bias. Evidently, this discrepancy needs further
investigation both experimentally and via numerical simulations.

\section{Conclusions}

The physics required to construct an accurate model of the tSZ power
spectrum is complicated. The gas pressure profiles at $r \simgt
r_{500}$ depend on an accurate modeling of non-thermal motions in the
intra-cluster medium. The profiles on smaller scales require an
accurate model of star formation and various feedback
processes. Furthermore, since the amplitude of the tSZ effect is
independent of redshift, these processes need to be modeled accurately
to high redshift ($z \simgt 1$) to predict the power spectrum at multipoles
$\ell \simgt 1000$.

In this paper, we have presented a simple model for the tSZ power
spectrum that is based on the X-ray inferred `universal' gas pressure
profile of Arnaud \etals (2010) extrapolated to higher redshift.  The
model is consistent with the low amplitude of the tSZ power spectrum
from recent observations from ACT and SPT and is consistent with
recent Planck observations of the tSZ effect for X-ray clusters with
redshifts $z \simlt 1$. Our model suggests that the `universal'
pressure profile extrapolated assuming nearly self-similar evolution,
provides an acceptable description of the observations.

We have shown that our model provides good fits to the tSZ power
spectra from recent numerical simulations, to an accuracy that is well
within the theoretical uncertainties involved in such simulations. Our
model may therefore be useful as a simple tSZ template,  since it has
only two key parameters (the overall amplitude and the evolution
parameter $\epsilon$) and is empirically motivated.

\medskip

\noindent
{\bf Acknowledgments:} We thank Eiichiro Komatsu for useful
correspondence concerning the KS02 paper. MM acknowledges 
support from ASI through ASI/INAF Agreement I/072/09/0 for Planck 
LFI Activity of Phase E2. GPE is supported by grants from STFC 
and  UKSA.


\begin{thebibliography}{}


\bibitem[\protect\citename{Arnaud \etals 2010}]{Arnaud10}
Arnaud M., Pratt G.W., Piffaretti R., B\"ohringer H., 
Croston J.H., Pointecouteau E., 2010, A\&A, 517, 92.

\bibitem[\protect\citename{ Battaglia \etals 2010}]{Battaglia10}
Battaglia N., Bond J.R., Pfrommer C., Sievers J.L, Sijacki D.,
2010, ApJ, 725, 91.

\bibitem[\protect\citename{ Bode, Ostriker and Vikhlinin 2009}]{Bode09}
Bode P., Ostriker J.P., Vikhlinin A., 2009, ApJ, 700, 989.

\bibitem[\protect\citename{ Bond and Myers 1996}]{BM06}
Bond J.R., Myers S., 1996, ApJS, 103, 1.

\bibitem[\protect\citename{ Bond \etals 2005}]{Betal05}
Bond J.R. \etal, 2005,  ApJ, 626, 12.

\bibitem[\protect\citename{Carlstrom \etals 2002}]{Carlstrom02}
Carlstrom J. E., Holder G. P., Reese E. D., 2002, ARAA, 40, 643.

\bibitem[\protect\citename{Cole and Kaiser 1988}]{CK98}
Cole S., Kaiser N., 1988, MNRAS, 233, 637.

\bibitem[\protect\citename{Cooray 2000}]{Cooray00}
Cooray A., 2000, PRD, 62, 3506.

\bibitem[\protect\citename{Cooray 2001}]{Cooray01}
Cooray A., 2001, PRD, 64, 3516.

\bibitem[\protect\citename{da Silva \etals 2000}]{daS00}
da Silva A.C., Barbosa D., Liddle A.R., Thomas P.A., 2000, MNRAS,
317, 37.


\bibitem[\protect\citename{Duffy \etals 2008}]{Duffy08}
Duffy, A.R., Battye R.A., Davies R.D., Moss A., Wilkinson P.N., 2008, MNRAS, 
383, 150.

\bibitem[\protect\citename{Dunkley \etals 2010}]{Dunkley10}
Dunkley J. \etal, 2010,  arXiv/astroph:1009.0866.

\bibitem[\protect\citename{Jenkins \etals 2001}]{Jenkins01}
Jenkins A.,  Frenk C. S., White S. D. M., Colberg J. M., 
 Cole S., Evrard A. E., Couchman H. M. P., Yoshida, N., 2001, MNRAS, 321, 372.

\bibitem[\protect\citename{Iliev \etals 2007}]{Iliev07}
Iliev I. T., Pen U.,  Bond J. R.,  Mellema G.,  Shapiro P. R.
2007, ApJ., 660, 933.

\bibitem[\protect\citename{Komatsu and Seljak 2001}]{KS01}
Komatsu E., Seljak U., 2001, MNRAS,  327, 1353.

\bibitem[\protect\citename{Komatsu and Seljak 2002}]{KS02}
Komatsu E., Seljak U., 2002, MNRAS,  336, 1256.

\bibitem[\protect\citename{Komatsu \etals 2011}]{Komatsu11}
Komatsu E. \etal, 2011, ApJS, 192, 18.

\bibitem[\protect\citename{Lau, Kratsov and Nagai 2002}]{Lau09}
Lau E. T., Kravtsov A. V., Nagai D., 2009, ApJ, 705, 1129.


\bibitem[\protect\citename{Lueker \etals 2010 }]{Lueker10}
Lueker M. \etals 2010, ApJ,  719, 1045.

\bibitem[\protect\citename{Melin, Bartlett and Delabrouille 2006}]{Melin06}
Melin J., Bartlett J.G.,  Delabrouille J., 2006, A\&A, 459, 341.

\bibitem[\protect\citename{Nagai 2011}]{N11}
 Nagai D., 2011, arXiv:1101.1322.

\bibitem[\protect\citename{Nagai, Vikhlinin and Kravtsov 2007}]{NVK07}
 Nagai D., Vikhlinin A., Kravtsov A.V., 2007, ApJ, 668, 1.

\bibitem[\protect\citename{Navarro, Frenk and White 1997}]{NFW97}
Navarro J.F., Frenk C.S., White S.D.M., 1997, ApJ, 490, 493.

\bibitem[\protect\citename{Piffareti \etals 2011}]{P11}
Piffareti R., Arnaud M., Pratt G.W., Pointecouteau E., Melin J.,
2011, in preparation.



\bibitem[\protect\citename{Planck Collaboration 2011a}]{Planck11a}
Planck Collaboration 2011a, arXiv:1101.2022.

\bibitem[\protect\citename{Planck Collaboration 2011b}]{Planck11b}
Planck Collaboration 2011b, arXiv:1101.2024.

\bibitem[\protect\citename{Planck Collaboration 2011c}]{Planck11c}
Planck Collaboration 2011c, arXiv:1101.2026.

\bibitem[\protect\citename{Planck Collaboration 2011d}]{Planck11d}
Planck Collaboration 2011d, arXiv:1101.2043.

\bibitem[\protect\citename{Planck Collaboration 2011e}]{Planck11e}
Planck Collaboration 2011e, arXiv:1101.2027.

\bibitem[\protect\citename{Plagge \etals 2010}]{Plagge10}
Plagge, T \etals 2010, ApJ, 716, 1118.

\bibitem[\protect\citename{Press and Schechter 1974}]{PS74}
Press W.H., Schechter P., 1974, ApJ,  187, 425.

\bibitem[\protect\citename{Sehgal \etals 2010}]{Sehgal10}
Sehgal N.,  Bode P.,  Das S.,  Hernandez-Monteagudo C., 
Huffenberger K.,  Lin Yen-Ting,  Ostriker J. P.,  Trac H., 2010, ApJ, 709, 920.


\bibitem[\protect\citename{Shaw \etals 2010 }]{Shaw10}
Shaw L.D., Nagai D., Bhattacharya S., Lau E.T.,  2010, ApJ,  725, 1425.

\bibitem[\protect\citename{Simionescu \etals 2011}]{Simionescu11}
Simionescu A. \etals 2011, Science, 331, 1576.

\bibitem[\protect\citename{Springel, White and Hernquist  2001 }]{SWH01}
Springel V., White M, Hernquist L., 2001, ApJ, 549, 681.

\bibitem[\protect\citename{Sun \etals 2011}]{Sun11}
Sun M. \etals 2011, ApJL, 727, 49.

\bibitem[\protect\citename{Sunyaev and  Zeldovich 1972}]{Sunyaev72}
Sunyaev R. A.,  Zeldovich Ya. B., Commments on Astrophysics and Space
Science, 20, 189.

\bibitem[\protect\citename{Trac \etals 2011}]{Trac11}
Trac H., Bode P., Ostriker J.P., 2011, ApJ,  727, 94.

\bibitem[\protect\citename{Urban \etals 2011}]{Urban11}
Urban O., Werner N., Simionescu A., Allen S. W., B\"ohringer H., 2011, arXiv:1102.2430.

\end{thebibliography}
\end{document}